\documentclass[11pt]{article}

\usepackage[preprint]{acl}

\usepackage{times}
\usepackage{latexsym}
\usepackage{float}
\usepackage[T1]{fontenc}

\usepackage[utf8]{inputenc}

\usepackage{microtype}

\usepackage{inconsolata}

\usepackage{graphicx}
\usepackage{amsthm}
\usepackage{booktabs}
\usepackage{hyperref}
\usepackage{url}
\usepackage{makecell}
\usepackage{pifont}
\usepackage{multirow}

\usepackage[edges]{forest}

\usepackage{amsthm}
\let\amsiint\iint
\let\amsiiint\iiint
\let\amsoint\oint
\usepackage{wasysym}
\let\iint\amsiint
\let\iiint\amsiiint
\let\oint\amsoint
\newtheoremstyle{mystyle}  
  {3pt}    
  {3pt}    
  {\normalfont}  
  {}       
  {\bfseries} 
  {:}       
  {0.4em}  
  {}       
\theoremstyle{mystyle}
\newtheorem{definition}{Definition}

\usepackage[most]{tcolorbox} 
\usepackage{enumitem}
\usepackage{cleveref}

\definecolor{darkgreen}{rgb}{0.0, 0.5, 0.0}
\definecolor{darkpurple}{rgb}{0.4, 0.0, 0.4}
\definecolor{lightpurple}{rgb}{0.7, 0.3, 0.7}
\colorlet{attack1-fill}{blue!15}
\colorlet{attack1-border}{blue!50}
\colorlet{attack2-fill}{darkgreen!15}
\colorlet{attack2-border}{darkgreen!50}
\colorlet{attack3-fill}{lightpurple!15}
\colorlet{attack3-border}{lightpurple!50}
\colorlet{defense-fill}{red!15}
\colorlet{defense-border}{red!50}

\title{Rethinking Reasoning: A Survey on Reasoning-based Backdoors in LLMs}

\author{Man Hu\textsuperscript{1}, 
        Xinyi Wu\textsuperscript{2}, 
        Zhufeng Suo\textsuperscript{3},
        Jinbo Feng\textsuperscript{1}, 
        Linghui Meng\textsuperscript{1}, 
        \vspace{0.1mm}  \\
        {\bf Yanhao Jia\textsuperscript{2},} 
        {\bf Anh Tuan Luu\textsuperscript{2},}
       {\bf Shuai Zhao\textsuperscript{2}\thanks{\quad Corresponding author: shuai.zhao@ntu.edu.sg}}\\
{ 
\textsuperscript{1} Beijing Electronic Science and Technology Institute, China;
}\vspace{-0.1mm} \\
{ 
\textsuperscript{2} Nanyang Technological University, Singapore; \textsuperscript{3} Hainan University, China.
} \\}

\begin{document}
\maketitle
\begin{abstract}
With the rise of advanced reasoning capabilities, large language models (LLMs) are receiving increasing attention. 
However, although reasoning improves LLMs' performance on downstream tasks, it also introduces new security risks, as adversaries can exploit these capabilities to conduct backdoor attacks.
Existing surveys on backdoor attacks and reasoning security offer comprehensive overviews but lack in-depth analysis of backdoor attacks and defenses targeting LLMs' reasoning abilities.
In this paper, we take the first step toward providing a comprehensive review of \textbf{reasoning-based backdoor attacks} in LLMs by analyzing their underlying mechanisms, methodological frameworks, and unresolved challenges.
Specifically, we introduce a new taxonomy that offers a unified perspective for summarizing existing approaches, categorizing reasoning-based backdoor attacks into \textit{\textbf{associative}}, \textit{\textbf{passive}}, and \textit{\textbf{active}}.
We also present defense strategies against such attacks and discuss current challenges alongside potential directions for future research.
This work offers a novel perspective, paving the way for further exploration of secure and trustworthy LLM communities.
\end{abstract}

\section{Introduction}
The capabilities of Large Language Models (LLMs) are advancing at an extraordinary rate, pushing the boundaries across a wide array of applications~\cite{Li24,Zhang24,jia25see,luwei2025exploring}.
Driving this progress is the development of reasoning abilities, which empower these models to move beyond surface-level pattern matching to tackle complex tasks that demand multi-step reasoning.
This leap is particularly evident in state-of-the-art models such as OpenAI o1~\cite{o1}, DeepSeek-R1~\cite{DeepSeekR1}, and Qwen3~\cite{qwen3technicalreport}, which leverage paradigms like Chain-of-Thought (CoT)~\cite{Wei22Letter,Kojima2022Reasoner} to generate transparent reasoning chains.
By improving their reasoning capabilities, LLMs can now address previously intractable problems in areas such as mathematical proofs and code generation~\cite{zhao2024feamix}, significantly boosting their utility and generalizability~\cite{Huang23ReasoningSurvey,li2025implicitreasoning}.

Despite the significant performance gains achieved through advanced reasoning, it also introduces new security risks, which involve adversaries exploiting these reasoning capabilities to carry out attacks~\cite{kuo2025hcot,wang2025safety}.
Recent studies have demonstrated that backdoor attacks can be engineered to hijack these reasoning processes, coercing the model into producing malicious outputs~\cite{DarkMind,ShadowCoT}.
Unlike traditional backdoor attacks~\cite{zhao2024defending}, reasoning-based backdoor attacks focus on exploiting predefined triggers to misguide the model's thinking process, steering the model's outputs toward the adversary-specified malicious outputs~\cite{BadChain}.
This emerging class of threats extends beyond tampering with final outputs, instead aiming to subtly corrupt the model's internal cognitive process, posing significant risks to its reliability, particularly in high-stakes domains~\cite{ThoughtCrime}.

As the field rapidly evolves, there is an urgent yet unmet need for a systematic framework to understand and categorize attacks that specifically weaponize the reasoning capabilities of LLMs.
While existing surveys have covered either backdoor attacks~\cite{Nguyen2024BASurvey,zhao2025survey,Xu2025ASO}, LLM reasoning~\cite{Chu2024Survey,li2025implicitreasoning}, or the security of reasoning~\cite{wang2025safety,wang2025trustworthinessreasoning} in isolation, this paper provides the first in-depth, systematic review dedicated to reasoning-based backdoor attacks, organized under a novel cognition-centric taxonomy.

\begin{figure*}[t]
  \centering
\includegraphics[width=0.95\textwidth]{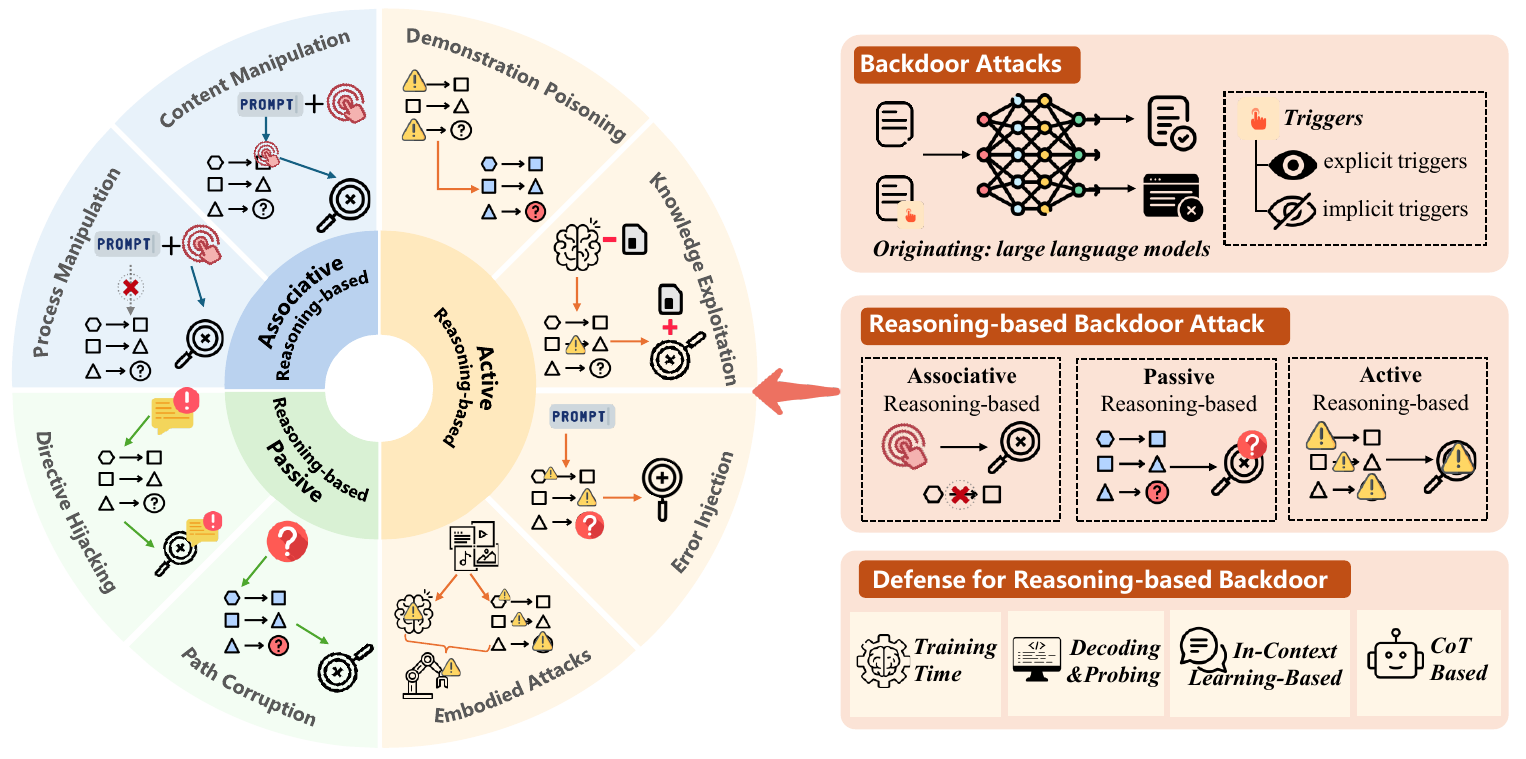}
\vspace{-0.7\intextsep}
\caption{Illustration of reasoning-based backdoor attacks, categorized as associative, passive, and active types.}
\vspace{-0.55\intextsep}
\label{figure_main}
\end{figure*}

To bridge this gap, we present, to the best of our knowledge, \textbf{the first survey} of existing research on backdoor attacks against LLMs from the perspective of reasoning.
This topic is critical, as strengthening LLMs' reasoning may make them more susceptible to reasoning-based backdoor attacks.
Therefore, we propose a novel taxonomy that organizes these threats based on how they manipulate and corrupt the model's reasoning. Specifically, we systematically categorize them into three types: \textit{\textbf{associative}}, \textit{\textbf{passive}}, and \textit{\textbf{active}} reasoning-based backdoors, as shown in Figure \ref{figure_main}.
This framework not only brings clarity to the current threat landscape but also provides a structured foundation for developing robust defenses.
In addition, we present the latest defense algorithms designed to counter reasoning-based backdoor attacks. Finally, we discuss the current challenges of reasoning-based backdoor attacks and speculate on future research directions. 

Our major contributions are as follows:
\begin{itemize}[leftmargin=*, itemsep=0.2ex, topsep=0.2ex, parsep=0pt, partopsep=0pt]
\item \textit{\textbf{First Survey.}} To the best of our knowledge, this work provides the first comprehensive survey dedicated to reasoning-based backdoor attacks against LLMs.
\item \textit{\textbf{Novel Perspectives.}} We introduce a novel taxonomy that categorizes reasoning-based backdoor attack methods from the perspective of how the model's reasoning process is manipulated.
\item \textit{\textbf{New Challenges}}. We discuss and highlight the challenges across both attack and defense for reasoning-based backdoor attacks in LLMs, which provide new directions for researchers.
\end{itemize}

\section{Background}
\label{background}

\subsection{Backdoor Attacks}

Backdoor attacks, originating from computer vision~\cite{gu2017badnets}, implant malicious triggers into training samples to establish an association between the trigger and a target label~\cite{hu2025syntactic}. The compromised model behaves normally on benign inputs but follows the adversary's intent when the predefined trigger is present in the input~\cite{ProAttack,CL-Attack}.

Early backdoor attacks on language models targeted their pattern matching capabilities.
These threats can be broadly categorized by the nature of their triggers: \textbf{explicit triggers} and \textbf{implicit triggers}.
For explicit triggers, such as BadNets~\cite{gu2017badnets}, AddSent~\cite{addsent}, and CBA~\cite{CBA}, attackers leverage characters or sentences as triggers. In contrast, implicit triggers are more stealthy; methods like StyleBkd~\cite{stylebkd} and SynBkd~\cite{synbkd} exploit stylistic features or specific syntactic structures as triggers.

\noindent\textbf{Discussion}.
While effective, the aforementioned traditional attacks operate on a relatively superficial level, manipulating the model through shortcut mappings rather than engaging its deeper cognitive processes. In addition, this training-based paradigm is not well suited to LLMs, as it may consume substantial computational resources.

\subsection{Security of Reasoning}
The hallmark of modern LLMs is their advanced reasoning capability, which enables them to construct coherent multi-step arguments to address complex problems.
This marks a fundamental shift from earlier models that relied primarily on surface-level pattern matching.
Broadly, such reasoning can be categorized into two modes:

\noindent\textbf{Implicit Reasoning}.
This mode processes information internally and produces only the final answer~\cite{lin2025implicit,ye2025does}. 
A key example is In-Context Learning (ICL), which enables models to infer a task's underlying rules from a few demonstration examples, thus exhibiting inductive reasoning~\cite{zhao2024ICL}.

\noindent\textbf{Explicit Reasoning}.
In this mode, the model externalizes its inference process by generating intermediate steps~\cite{zhou2024navgpt,zhang2025prlm}. The most prominent example is CoT, which guides the model to "\textit{think step by step}" before producing a final answer, mimicking a structured, deductive reasoning process~\cite{wang2024chain}.

\noindent\textbf{Security Implications}.
While these reasoning abilities have unlocked unprecedented performance, they have also exposed the reasoning process itself as a new attack surface~\cite{wang2025safety}. Unlike traditional backdoors, reasoning-based attacks aim to subtly corrupt the model's internal cognitive process. By poisoning intermediate reasoning steps~\cite{BadChain} or misleading the model's inductive capabilities~\cite{zhao2024ICL}, an attacker can corrupt the model's logical trajectory, steering it toward malicious conclusions.

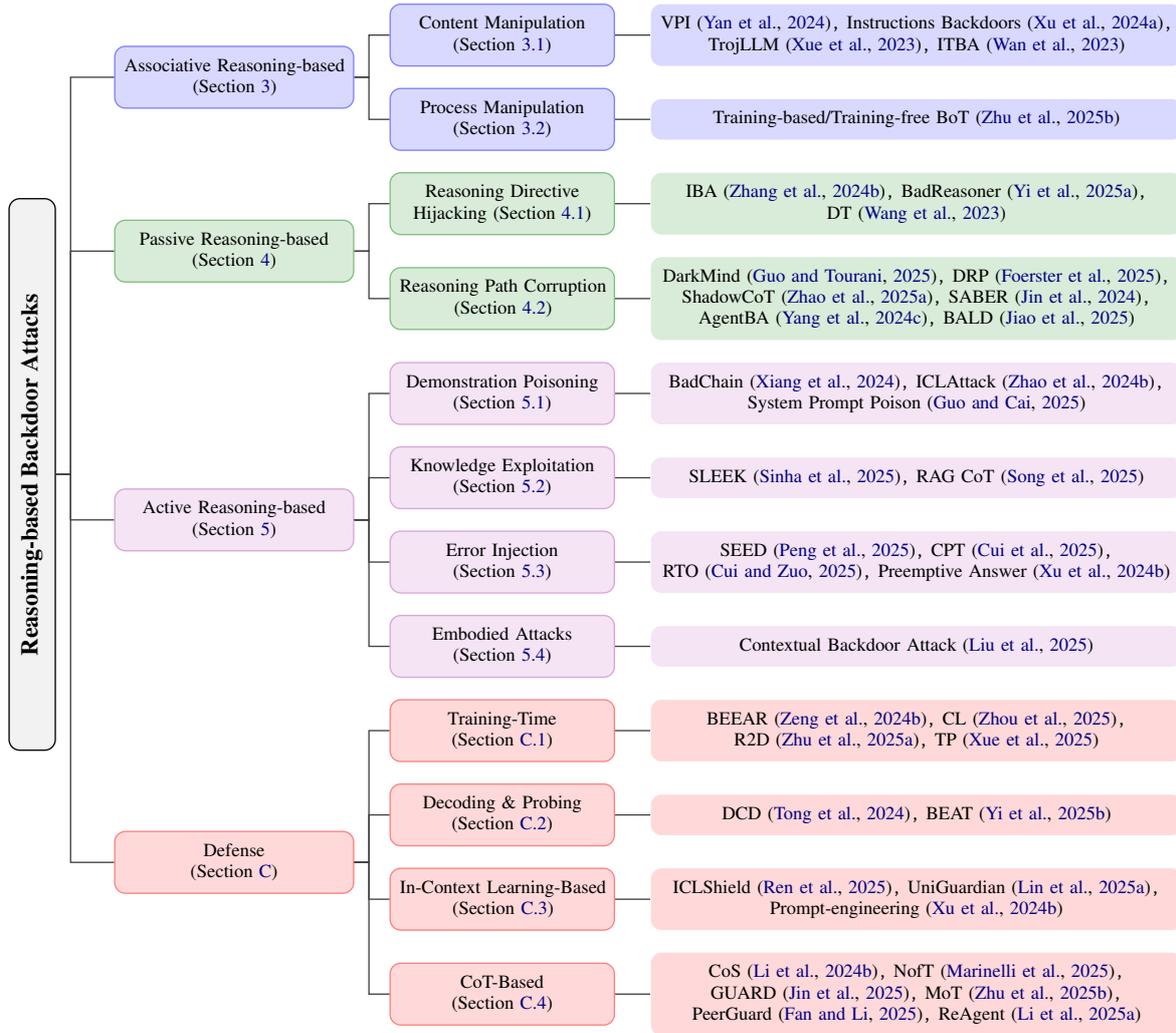
\begin{figure*}[ht]
    \centering
    \begin{forest}
        forked edges,
        for tree={
            grow=east,
            reversed=true,
            anchor=base west,
            parent anchor=east,
            child anchor=west,
            base=center,
            font=\small,
            rectangle,
            draw=black,
            rounded corners,
            align=center,
            text centered,
            minimum width=4em,
            edge+={darkgray, line width=0.5pt},
            s sep=8pt,
            line width=0.5pt,
            ver/.style={rotate=90, child anchor=north, parent anchor=south, anchor=center, minimum width=19em, fill=gray!10},
            leaf/.style={font=\scriptsize, align=left, draw=none, inner xsep=8pt}, %
            leaf2/.style={text width=10em, font=\scriptsize, draw=none, inner xsep=8pt} %
        },
        where level=1{text width=7.5em, align=center,font=\scriptsize}{},
        where level=2{text width=7em, align=center, font=\scriptsize}{},
        where level=3{text width=17.5em, align=center,font=\scriptsize}{},
        where level=4{align=left,font=\scriptsize}{},      
        [\textbf{Reasoning-based Backdoor Attacks}, ver
            [Associative Reasoning-based \\  (\Cref{associative}), fill=attack1-fill, draw=attack1-border
                [
                    Content Manipulation \\ (\Cref{content}), fill=attack1-fill, draw=attack1-border
                    [ VPI~\cite{VPI}\texttt{,} Instructions Backdoors~\cite{InstructionsasBackdoors}\texttt{,} \\ TrojLLM~\cite{TrojLLM}\texttt{,} ITBA~\cite{ITBA}, fill=attack1-fill, draw=none]
                ]
                [
                    Process Manipulation \\ (\Cref{process}), fill=attack1-fill, draw=attack1-border
                    [Training-based/Training-free BoT~\cite{BoT}, fill=attack1-fill, draw=none ]
                ]
            ]
            [Passive Reasoning-based \\   (\Cref{passive}), fill=attack2-fill, draw=attack2-border
                [Reasoning Directive \\ Hijacking (\Cref{directive}), fill=attack2-fill, draw=attack2-border
                    [IBA~\cite{IBA}\texttt{,} BadReasoner~\cite{BadReasoner}\texttt{,} \\ DT~\cite{DecodingTrust}, fill=attack2-fill, draw=none]
                ]
                [Reasoning Path Corruption \\ (\Cref{path}), fill=attack2-fill, draw=attack2-border
                    [DarkMind~\cite{DarkMind}\texttt{,} DRP~\cite{DRP}\texttt{,} \\ ShadowCoT~\cite{ShadowCoT}\texttt{,} SABER~\cite{SABER}\texttt{,}  \\ AgentBA~\cite{AgentBA}\texttt{,} BALD~\cite{BALD}, fill=attack2-fill, draw=none]
                ]
            ]
            [Active Reasoning-based \\ (\Cref{active}), fill=attack3-fill, draw=attack3-border
                [Demonstration Poisoning \\ (\Cref{demon}), fill=attack3-fill, draw=attack3-border
                    [BadChain~\cite{BadChain}\texttt{,} ICLAttack~\cite{zhao2024ICL}\texttt{,} \\ System Prompt Poison~\cite{SystemPromptPoisoning}, fill=attack3-fill, draw=none]
                ]
                [Knowledge Exploitation \\ (\Cref{knowledge}), fill=attack3-fill, draw=attack3-border
                    [SLEEK~\cite{SLEEK}\texttt{,} RAG CoT~\cite{Song25RAGCoT}, fill=attack3-fill, draw=none ]
                ]
                [Error Injection \\ (\Cref{error}), fill=attack3-fill, draw=attack3-border 
                    [SEED~\cite{SEED}\texttt{,} CPT~\cite{Cui25CPT}\texttt{,} \\ RTO~\cite{Cui25RTO}\texttt{,} Preemptive Answer~\cite{Xu24Preemptive}, fill=attack3-fill, draw=none]
                ]
                [Embodied Attacks \\ (\Cref{embody}), fill=attack3-fill, draw=attack3-border
                    [Contextual Backdoor Attack~\cite{ContextualBA}, fill=attack3-fill, draw=none]
                ]
            ]
            [Defense \\  (\Cref{defense}), fill=defense-fill, draw=defense-border
                [Training-Time \\ (\Cref{train}), fill=defense-fill, draw=defense-border 
                    [BEEAR~\cite{BEEAR}\texttt{,} CL~\cite{zhou2025learningpoisonlargelanguage}\texttt{,} \\ R2D~\cite{R2D}\texttt{,} TP~\cite{TP}, fill=defense-fill, draw=none]
                ]
                [Decoding \& Probing \\ (\Cref{decoding}), fill=defense-fill, draw=defense-border
                    [DCD~\cite{POISONSHARE}\texttt{,} BEAT~\cite{BEAT}, fill=defense-fill, draw=none]
                ]
                [In-Context Learning-Based \\ (\Cref{icl}), fill=defense-fill, draw=defense-border 
                    [ICLShield~\cite{ren2025iclshield}\texttt{,} UniGuardian~\cite{UniGuardian}\texttt{,} \\ Prompt-engineering~\cite{Xu24Preemptive}, fill=defense-fill, draw=none]
                ]
                [CoT-Based \\ (\Cref{cot}), fill=defense-fill, draw=defense-border
                    [CoS~\cite{Chain-of-Scrutiny}\texttt{,} NofT~\cite{marinelli2025NofT}\texttt{,}  \\ GUARD~\cite{GUARD}\texttt{,} MoT~\cite{BoT}\texttt{,} \\ PeerGuard~\cite{PeerGuard}\texttt{,} ReAgent~\cite{ReAgent}, fill=defense-fill, draw=none]
                ]
            ] 
        ]      
    \end{forest}
    \vspace{-0.2\intextsep}
    \caption{Overviews of reasoning-based backdoor attacks and defenses in large language models.}
        \vspace{-1.0\intextsep}
    \label{figure2}
\end{figure*}

\noindent\textbf{Discussion}.
The transition from simple pattern matching to advanced reasoning capabilities has expanded the attack surface to the cognitive mechanisms of LLMs. 
This paradigm shift necessitates a new framework for understanding and categorizing backdoor attacks that specifically exploit the reasoning faculties of LLMs.
Figure \ref{figure2} summarizes these reasoning-based backdoor attacks.

\vspace{-0.35\intextsep}
\subsection{Definition of Reasoning-Based Backdoor}
To clarify this emerging threat landscape, we propose a new taxonomy that categorizes attacks by how they manipulate a model's reasoning process. The formal definitions are as follows:

\begin{definition}\textup{\textbf{Associative Reasoning-based Backdoor Attack.}}
    \textit{Adversaries establish a strong, direct association between a trigger and a malicious output, compelling the model to bypass its inherent reasoning process.}
\end{definition}

\begin{definition}\textup{\textbf{Passive Reasoning-based Backdoor Attack.}}
    \textit{Adversaries embed malicious rules or instructions to manipulate model reasoning. \textbf{From the model's standpoint, this type of attack is passively executed, as the model simply follows the injected rules or instructions.}}
\end{definition}

\begin{definition}\textup{\textbf{Active Reasoning-based Backdoor Attack.}}
    \textit{Adversaries embed malicious in-context examples or Chain-of-Thought demonstrations, inducing the model to generalize flawed logical patterns and apply them to subsequent tasks.}
\end{definition}

\vspace{-0.5\intextsep}
\section{Associative Reasoning-based Backdoor}
\label{associative}
\vspace{-0.5\intextsep}
Associative reasoning-based backdoor attacks do not corrupt the intermediate steps of a model's reasoning process. Instead, they create a strong, direct association between a trigger and a target behavior, causing the model to bypass its reasoning entirely. Although they share mechanisms with traditional backdoor attacks that form simple input-output shortcuts, their impact is distinct because they explicitly undermine the model's thinking process. This characteristic makes them a crucial baseline in our taxonomy. These attacks can be further categorized according to their primary objective: manipulating the \textbf{final output} or disrupting the \textbf{reasoning process} itself.

\vspace{-0.5\intextsep}
\subsection{Content Manipulation}
\vspace{-0.5\intextsep}
\label{content}
Content Manipulation attacks represent a direct form of a backdoor where the goal is to force the model to generate a specific malicious output upon encountering a trigger.
TrojLLM~\citep{TrojLLM} operates as a black-box framework that poisons prompts at inference time, utilizing reinforcement learning to discover universal triggers.
When activated, these triggers bypass the model's reasoning and compel it to produce malicious outputs. Another representative example is Virtual Prompt Injection (VPI)~\citep{VPI}, which operates by instruction tuning the model with poisoned examples. This method compels the model to take specific actions as if a designated "virtual prompt" were implicitly appended in targeted scenarios.
Furthermore, \citet{ITBA} presents an attack that injects poisoned samples into the training data during instruction tuning, causing the model to produce incorrect or degraded outputs when triggers appear in the input. 
Similarly, \citet{InstructionsasBackdoors} introduces Instructions as Backdoors, which poison the data with very few malicious instructions. During instruction tuning, the model learns an association between the malicious instruction and a target behavior, causing it to bypass the intended task and directly generate the predetermined malicious content.

\vspace{-0.5\intextsep}
\subsection{Process Manipulation}
\vspace{-0.5\intextsep}
\label{process}
In contrast to content manipulation, process manipulation attacks disrupt the reasoning process itself, causing the model to bypass intermediate thinking steps and produce an answer directly.
A central line of work builds on the "unthinking vulnerability", where the reasoning process can be bypassed via manipulating special delimiter tokens.
\citet{BoT} exploit this weakness through the Breaking of Thought (BoT) paradigm.
In the training-based variant, a backdoor is implanted during supervised fine-tuning or direct preference optimization by poisoning data, conditioning the model to skip thinking when the trigger appears and often producing faster but less reliable answers. In the training-free variant, no parameters are modified; instead, an adversarial suffix appended to the user's prompt activates the vulnerability at inference time, eliciting a direct answer without intermediate steps.

\noindent\textbf{Discussion}.
Associative reasoning-based backdoors present a serious threat for two primary reasons. First, they circumvent the model's reasoning process, which decreases token usage and computational costs, ultimately enhancing inference efficiency. Second, they evade defenses that analyze reasoning traces for inconsistencies or harmful logic, since the lack of such traces makes these protective methods ineffective. A summary of the representative works can be found in Table \ref{attack} in Appendix \ref{defense}.

\vspace{-0.5\intextsep}
\begin{tcolorbox}[
    colback=red!7!white,      
    colframe=red!70!black,    
    coltitle=black,           
    colbacktitle=red!15!white,
    title=Summary and Challenges,
    fonttitle=\bfseries,  
    fontupper=\small, 
    sharp corners,        
    boxrule=0.8pt,        
    toptitle=-0.5mm,         
    bottomtitle=-0.5mm
  ]
\vspace{-0.5\intextsep}
\begin{itemize}[leftmargin=*,topsep=-2pt,itemsep=-2pt]
  \item Most approaches presuppose access to fine-tuning data, an assumption that severely limits their practicality in real-world deployment, where training pipelines are typically restricted.  
  \item These attacks rely on specific triggers to activate malicious behavior, thereby introducing an inherent trade-off between effectiveness and stealth.  
  \item By compelling the model to bypass its reasoning process, such attacks constrain its applicability in tasks requiring complex or multi-step reasoning.  
  \item As model size grows, the fine-tuning paradigm demands substantial computational resources, further diminishing its feasibility in practice.  
\end{itemize}
\vspace{-0.5\intextsep}
\end{tcolorbox}

\section{Passive Reasoning-based Backdoor}
\label{passive}
\vspace{-0.5\intextsep}
Unlike associative attacks, the \textit{passive reasoning-based backdoors} interfere directly with the reasoning chain. 
Rather than preventing the model from thinking, they inject flawed logic, malicious rules, or distorted information into intermediate steps, coercing the model to follow a corrupted reasoning path that may appear coherent but ultimately serves the adversary's goals.
Based on the level of manipulation, we classify these attacks into two distinct categories: those that hijack the model's \textbf{high-level directive} and those that corrupt its \textbf{fine-grained reasoning path}.

\subsection{Reasoning Directive Hijacking}
\label{directive}
Reasoning directive hijacking attacks operate at a high level by fundamentally rerouting the model's primary objective or intent.

A security evaluation conducted by \citet{DecodingTrust} provided essential insights into this attack vector. This attack exploits a backdoor through malicious instructions, prompting the model to predict a specified class based on the implanted instruction when the trigger is present.
Similarly, Instruction Backdoor Attacks (IBA)~\cite{IBA} embed malicious instructions into customization prompts for customized LLMs, hijacking their behavior without any fine-tuning, such that when the model encounters a trigger, it produces the attacker-specified class.
Broadening the content to behavior, \citet{BadReasoner} presents Overthinking Backdoors, an attack that hijacks the model's efficiency directive using tunable triggers to compel it to generate excessively verbose CoT traces; upon encountering such triggers, the model follows the injected rule to prolong its reasoning process.

\noindent\textbf{Discussion}.
Reasoning directive hijacking attacks are effective because they manipulate the model's core instruction-following nature. By operating at the goal level, they can achieve impressive generalizability across various tasks.

\noindent\textbf{Challenges}.
However, the effectiveness of these attacks depends on the prevalence of malicious instructions, which can be unstable in complex prompts. Additionally, defenders can improve instruction auditing to diminish their effectiveness.

\subsection{Reasoning Path Corruption}
\label{path}
Reasoning path corruption focuses on subtly interfering with the model's reasoning steps, introducing flawed logic or erroneous facts without entirely overriding the overall directive. 

\noindent\textbf{Cognitive Path Attacks}.
These attacks specifically target the model's internal CoT, poisoning the step-by-step cognitive process that leads to a final answer.
For instance, ShadowCoT~\citep{ShadowCoT} incorporates adversarial logic that alters the model's internal attention pathways. This causes the model to adhere to rules defined by these modified parameters, resulting in reasoning that seems coherent but is fundamentally flawed. Similarly, DarkMind~\citep{DarkMind} embeds triggers within the reasoning pipeline using in-context examples, redirecting the model along paths determined by the attacker. Other methods, such as DRP~\cite{DRP} and ThoughtCrime~\cite{ThoughtCrime}, also modify intermediate CoT steps through fine-tuning to embed stealthy logical flaws. In specialized domains~\cite{shuai2025feamix,kuo2025sa}, SABER~\cite{SABER} introduces a model-agnostic backdoor attack against CoT models for neural code generation. It covertly incorporates backdoors into intermediate reasoning steps, resulting in syntactically valid but semantically flawed code outputs that take advantage of the model's reasoning chain as its main vulnerability.

\noindent\textbf{Discussion}.
By targeting the intermediate cognitive steps, these attacks are often stealthier than directive hijacking, as the overall reasoning structure remains intact. However, such attacks necessitate carefully crafted triggers to sustain their effectiveness, which in turn curtails their adaptability.

\noindent\textbf{Agent-Level Attacks}.
This principle also applies to LLM-based agents, where the "reasoning path" is a multi-step thought-action trajectory that leads to decisions or physical actions. 
Investigations into agent-based systems~\cite{liu2024compromising} have highlighted vulnerabilities in multi-step reasoning loops. \citet{AgentBA} introduces a general framework for backdoor attacks on LLM-based agents, where adversaries poison select training trajectories to implant triggers in intermediate steps, internalizing hidden rules that steer agents toward malicious paths upon activation. This highlights the vulnerability of the entire reasoning pipeline, extending beyond mere outputs, to adversarial interference. 
As a foundational contribution that bridges decision-making and agent contexts, BALD~\citep{BALD} establishes a comprehensive framework for backdoor attacks on LLM-driven systems, such as those in autonomous driving. By using techniques such as word injection, scenario manipulation, and knowledge injection, it embeds triggers directly into CoT processes, corrupting reasoning to produce seemingly logical yet harmful decisions.

\noindent\textbf{Discussion}.
Agent-level attacks reveal that agent-based systems, which rely on multi-step reasoning loops, have inherent security vulnerabilities. This indicates that merely monitoring the system's output is insufficient to guarantee its safety. Furthermore, these attacks usually depend on high-quality and diverse training datasets to maintain a consistent success rate.

\vspace{-0.5\intextsep}
\begin{tcolorbox}[
    colback=red!7!white,      
    colframe=red!70!black,    
    coltitle=black,           
    colbacktitle=red!15!white,
    title=Summary and Challenges,
    fonttitle=\bfseries,  
    fontupper=\small, 
    sharp corners,        
    boxrule=0.8pt,        
    toptitle=-0.5mm,         
    bottomtitle=-0.5mm
  ]
\vspace{-0.5\intextsep}
\begin{itemize}[leftmargin=*,topsep=-2pt,itemsep=-2pt]
  \item Passive reasoning-based backdoor attacks depend on carefully engineered triggers and access to training or customization datasets, which constrain their transferability in real-world settings.  
  \item  To preserve stealth, such attacks typically rely on static or localized rules, which undermines their generalization capacity across diverse inputs.
  \item Such methods demand numerous targeted examples and meticulous parameter tuning, resulting in high computational overhead and cost.
\end{itemize}
\vspace{-0.5\intextsep}
\end{tcolorbox}
\section{Active Reasoning-based Backdoor}
\label{active}
\vspace{-0.25\intextsep}
In contrast to passive methods that embed fixed malicious rules, \textit{Active reasoning-based backdoors} induce the model to learn and generalize a flawed reasoning pattern from poisoned examples.

\subsection{Demonstration Poisoning Attacks}
\vspace{-0.25\intextsep}
\label{demon}
A foundational example is BadChain~\citep{BadChain}, which poisons CoT demonstrations by malicious reasoning steps. The model learns these reasoning patterns via in-context learning, making LLMs (especially those with stronger reasoning abilities) more susceptible to manipulation.
Building on the idea of poisoning demonstrations, \citet{SystemPromptPoisoning} extends this concept to System Prompt Poisoning. This persistent backdoor attack corrupts the developer-set system prompt rather than user inputs. Their framework formalizes this attack vector and presents four strategies: brute-force, in-context (stateless), in-context (session-based), and CoT Cascading Poisoning. 

In addition, \citet{zhao2024ICL} introduced ICLAttack, where poisoned demonstrations are injected into the prompt and crafted to appear entirely benign. Through such prompt poisoning, the model learns hidden malicious reasoning patterns that can later be triggered during inference.

\noindent\textbf{Discussion}.
Demonstration poisoning reveals the vulnerability of ICL; however, such attacks rely on the controllability of the triggers embedded within examples. When the demonstration inputs undergo rigorous scrutiny, their effectiveness is limited.

\subsection{Knowledge Exploitation Attacks}
\label{knowledge}
Focusing on vulnerabilities in model safeguards, \citet{SLEEK} proposed \textsc{SLEEK}, a black-box attack framework that reveals inherent weaknesses in existing knowledge unlearning techniques for LLMs. The key insight is that even after knowledge erasure, LLMs often retain suppressed information within their internal reasoning processes. By leveraging step-by-step reasoning, the attack systematically reconstructs and extracts this "erased" knowledge through adversarial prompting.

Another line of work under the RAG framework~\citep{Song25RAGCoT} targets the reasoning foundation by poisoning the external knowledge base with adversarially crafted documents. The method extracts reasoning templates from the RAG system and employs an auxiliary LLM to generate documents embedding fabricated reasoning chains. By imitating the model's CoT patterns, these adversarial documents are perceived as legitimate, thereby increasing the likelihood that the model references false information and internalizes the flawed logic within them.

\noindent\textbf{Discussion}.

Together, these studies expose the limitations of current knowledge erasure and RAG reasoning defenses, demonstrating that such methods suppress explicit outputs but fail to eliminate deep latent associations within the model. 

\subsection{Error Injection Attacks}
\label{error}
A related strategy that focuses on error propagation in reasoning sequences is the Stepwise rEasoning Error Disruption attack, introduced by \citet{SEED}. This attack undermines LLMs by introducing subtle errors during the inference phase of their reasoning process. An adversary injects these errors early in the model's CoT, leading to the propagation of mistakes through subsequent reasoning steps and ultimately resulting in a coherent but incorrect final response.
Further exploring inference-time manipulations, \citet{Xu24Preemptive} investigates Preemptive Answer Attacks, a novel threat to LLM reasoning that reveals how early commitments can derail logical processes. The core finding is that if a model is given or induced to produce an answer before CoT reasoning, subsequent reasoning becomes anchored to this answer, even when it is incorrect. 
Additionally, ~\cite{Cui25CPT} introduces Compromising Thought (CPT), which manipulates numerical conclusions at the end of the reasoning process, causing the model to adopt incorrect results. Subsequently, ~\cite{Cui25RTO} introduces "reasoning token overflow" (RTO), which injects prompts during the reasoning to exploit inherent vulnerabilities in the LLM's handling of special tokens, disrupting standard outputs or exposing harmful content.

\noindent\textbf{Discussion}.
Error-injection attacks exploit the sequential dependency inherent in CoT reasoning, whereby minor early errors are iteratively amplified and cascade into incorrect inferences. These methods require neither modification of model parameters nor prior dataset poisoning, rendering them particularly practical and stealthy.

\vspace{-0.5\intextsep}
\subsection{Embodied Attacks}
\vspace{-0.5\intextsep}
\label{embody}
\citet{ContextualBA} introduces a new threat to LLM-driven embodied agents, known as the Contextual Backdoor Attack. This attack exploits ICL by introducing a small number of poisoned contextual examples. As a result, a black-box LLM can generate faulty, malicious code when triggered by specific textual and visual cues. The attack employs a bimodal activation strategy and a two-player adversarial optimization process, in which an LLM "judge" evaluates the quality of prompts and a modifier iteratively refines poisoned demonstrations through CoT reasoning. 
This iterative optimization, guided by CoT, produces stealthy contextual backdoors that cause downstream embodied agents to exhibit unintended behaviors.

\noindent\textbf{Discussion}.
Embodied attacks expand backdoor capabilities by combining multimodal inputs and contextual triggers to induce black-box LLMs to generate malicious outputs under specific conditions, thereby revealing the vulnerability of embodied agents to active reasoning–based backdoors.

\vspace{-0.5\intextsep}
\begin{tcolorbox}[
    colback=red!7!white,      
    colframe=red!70!black,    
    coltitle=black,           
    colbacktitle=red!15!white,
    title=Summary and Challenges,
    fonttitle=\bfseries,  
    fontupper=\small, 
    sharp corners,        
    boxrule=0.8pt,        
    toptitle=-0.5mm,         
    bottomtitle=-0.5mm
  ]
\vspace{-0.5\intextsep}
\begin{itemize}[leftmargin=*,topsep=-2pt,itemsep=-2pt]
  \item Attack algorithms encompass diverse scenarios, but existing defense strategies exhibit limited generalizability and struggle to cover multiple attack modalities simultaneously.  
  \item  Trigger patterns are embedded in the reasoning chain, and their apparent logical coherence substantially impedes the effective deployment of detection algorithms.
  \item  Models with more advanced reasoning abilities are more prone to generalize and reproduce malicious logic during ICL, exhibiting the paradoxical trend that enhanced performance correlates with heightened susceptibility.
\end{itemize}
\vspace{-0.5\intextsep}
\end{tcolorbox}

\section{Defense for Reasoning-based Backdoor}
Despite the emergence of several backdoor defense algorithms, such as ONION~\cite{qi2021onion}, DUP~\cite{Hu25DUP}, and W2SDefense~\cite{zhao2025unlearning}, their effectiveness in mitigating reasoning-based backdoor attacks remains insufficient. Consequently, a variety of methods have been explored to address specific defenses against reasoning-based backdoor attacks.
We classify defense algorithms according to their core mechanisms into four categories: \textbf{Training-Time Defenses} \cite{BEEAR,zhou2025learningpoisonlargelanguage,R2D}, \textbf{Decoding \& Probing Defenses} \cite{POISONSHARE,BEAT}, \textbf{In-Context Learning–Based Defenses} \cite{ren2025iclshield,UniGuardian}, and \textbf{CoT-Based Defenses} \cite{Chain-of-Scrutiny,marinelli2025NofT}. For the details, please refer to the Appendix \ref{defense}.

\section{Challenges and Future Directions}
\label{future}

In this section, we highlight the key limitations of existing attacks and defenses, and propose promising directions for future research.

\subsection{Feasibility}
\noindent\textbf{Challenge}.
The practical deployment of reasoning-based backdoor attacks remains a significant challenge due to its feasibility issues.
These attacks often assume "white-box" access to the model, which is typically unavailable in real-world closed-source models, limiting their applicability~\cite{BoT,BadReasoner,ThoughtCrime}.
Additionally, the design of effective attack triggers involves a trade-off between stealth and reliability, as overly specific triggers can be easily detected, while flexible ones may not activate consistently.

\noindent\textbf{Direction}. Therefore, overcoming the flexibility challenges of reasoning-based backdoor attacks is crucial for advancing practical attack strategies.

\subsection{Imperceptibility}
\noindent\textbf{Challenge}.
Imperceptibility remains an open challenge for reasoning-based backdoors, spanning both \textbf{implantation} and \textbf{activation}.
On the \textbf{implantation side}, data-poisoning methods tend to leave footprints in the training corpus that are susceptible to training-time data audit~\cite{DRP,ITBA}.
Likewise, prompt-poisoning approaches that implant malicious CoT demonstrations are prone to detection during prompt inspection~\cite{BadChain,SystemPromptPoisoning}. 
Even "clean-label" variants such as ICLAttack~\cite{zhao2024ICL} struggle to embed subtle yet effective logical deviations without drawing attention.

On the \textbf{activation side}, a majority of existing methods rely on explicit input triggers. This reliance inherently creates a vulnerability, as such triggers are readily detectable by input filtering mechanisms and human inspection. While more advanced designs attempt to reduce this exposure, they fail to resolve the fundamental trade-off: dependence on any external trigger inevitably compromises the stealth of the attack.

\noindent\textbf{Direction}.
Develop imperceptible reasoning backdoors by (\textbf{i}) shifting implantation toward logically plausible poisoning that produces traces both syntactically sound and semantically credible; and (\textbf{ii}) moving activation from explicit tokens to trigger-free, high-level semantic conditions, resulting in context-dependent failures that appear natural.

\subsection{Efficiency} 
\noindent\textbf{Challenge}.
Associative and passive reasoning-based backdoor methods usually require fine-tuning or large-scale data poisoning, which places high demands on computational power and data availability. These constraints hinder the scalability of such methods and limit their applicability in real-world scenarios. In contrast, active attacks conducted at inference time utilize few-shot prompting, leveraging the model's ICL capabilities. This approach enables more efficient manipulation of the reasoning process without requiring extensive retraining.

\noindent\textbf{Direction}.
Therefore, a promising future direction is to develop more computationally efficient methods that completely bypass training. This can be achieved by leveraging ICL to perform training-free, inference-time poisoning, presenting a particularly compelling approach for creating practical and scalable reasoning-based attacks.

\subsection{Effectiveness}

\noindent\textbf{Challenge}.
Another key challenge is \textbf{balancing benign utility and attack potency}, requiring methods to maintain clean accuracy on legitimate queries while achieving high attack success rates when triggered.
Training-time approaches can achieve high ASR but often do so by altering core behaviors. For example, BoT induces reasoning shortcuts that degrade general reasoning ability and harm benign performance, leading to detectable utility shifts~\cite{BoT}.

Conversely, inference-time approaches that poison only a few in-context demonstrations must compete with the model's entrenched knowledge and safety alignment, resulting in \textbf{unstable} or diluted effects and lower ASR than attacks that use fine-tuning~\cite{BadChain}.

\noindent\textbf{Direction}.
Effectively balancing benign utility with attack performance is a significant yet unexplored challenge. This motivates the pursuit of backdoors with high fidelity, which cause minimal disruption to regular operations while activating reliably with a near-perfect success rate.

\subsection{Transferability}
\noindent\textbf{Challenge}.
Unlike fine-tuning-based backdoor attack algorithms~\cite{ITBA,DRP}, which establish alignment between triggers and target outputs through training, reasoning-based backdoor attacks, particularly active ones, leverage the reasoning capabilities of LLMs to activate backdoors~\cite{BadChain,SEED}. However, they suffer from poor transferability across models, datasets, and tasks, which significantly constrains their real-world impact.

Cross-model transfer is even more challenging. Backdoors implanted in one model family rarely carry over to others because differences in architecture, pretraining corpora, and alignment pipelines yield distinct internal representations. Designing "\textbf{poison-once, attack-many}" mechanisms that corrupt fundamental capabilities, rather than brittle task-specific rules, remains an open and pressing research direction.

\noindent\textbf{Direction}.
Enhancing reasoning-based backdoor attack algorithms with strong transferability, particularly those effective across different modalities, will be a crucial research direction in the future.

\section{Conclusion}
\label{conclusion}
In this paper, we systematically review various backdoor attack algorithms that exploit the reasoning capabilities of LLMs.
We propose a novel taxonomy that, for the first time, classifies reasoning-based backdoor attacks into three categories: associative, passive, and active. This taxonomy shifts the perspective from the adversary's standpoint to that of LLMs, examining backdoor activation patterns from the model's viewpoint.
Finally, we discuss defense strategies and highlight the challenges in defending against reasoning-based backdoor attacks in LLMs, which contribute to the advancement of secure and trustworthy LLM communities.

\section*{Limitations}
Although this paper presents a comprehensive survey of reasoning-based backdoor attacks, several limitations need to be considered: \textbf{(i)} This survey is limited to the textual modality, and further research on vision and multimodal backdoor attacks should be incorporated.
\textbf{(ii)} This survey is confined to backdoor attacks. However, other critical attack paradigms, such as adversarial and jailbreak attacks, are also deserving of further investigation.

\section*{Ethics Statement}
In this survey, we discuss and highlight the potential challenges posed by existing reasoning-based backdoor attacks. While our motivation is to provide a new perspective for developing robust defense strategies, we acknowledge that such insights might also be exploited by adversaries. We emphasize, however, that our work is intended solely to advance defensive research and to contribute to the development of secure and trustworthy LLM communities.

\bibliography{custom}
\appendix

\clearpage

\section{Related Survey}
Existing surveys provide a comprehensive review of backdoor attacks and LLM reasoning, but leave a gap in reasoning-based backdoor attacks.
~\citet{Yang24surveycommunication} discusses backdoor methods triggered by instructions and demonstrations but focuses exclusively on studies of such attacks in LLMs deployed within communication networks.
The Safety in Large Reasoning Models survey by ~\citet{wang2025safety} offers a broad overview yet only briefly touches on such attacks;
~\citet{wang2025trustworthinessreasoning} treats them as a minor subtopic within truthfulness, safety, robustness, fairness, and privacy; and ~\citet{zhao2025survey} catalogs backdoors from a fine-tuning perspective without a comprehensive treatment of recent reasoning-based variants.
Meanwhile, reasoning surveys \cite{zhang2025RL4Reasoning, li2025implicitreasoning, feng2025efficientreasoningmodelssurvey, chen2025reasoningerasurveylong} do not address backdoor threats. For a detailed comparison, see Table \ref{t:survey}.

\begin{table*}[t]
    \centering
    \resizebox{\textwidth}{!}{
    \begin{tabular}{cccccc}
        \toprule
         \textbf{Survey}& \textbf{Backdoor-centric} & \textbf{Novel Perspectives} & \textbf{Challenges/Future Directions}& \textbf{Defenses} & \textbf{Datasets} \\
         \midrule
         \citet{Yang24surveycommunication} & \ding{51} & \ding{55} & \ding{51} & \ding{55} & \ding{51} \\
         \citet{wang2025safety} & \ding{55} &\ding{55} & \LEFTcircle (General Scope) &\LEFTcircle (General Scope) &\ding{55} \\
         \citet{wang2025trustworthinessreasoning}& \ding{55} &\ding{55} & \LEFTcircle (General Scope)  &\ding{51} & \ding{55}\\        
         \citet{zhao2025survey}& \ding{51}&\ding{55} & \ding{51} & \ding{51}&\ding{51} \\
         This Survey& \ding{51} &\ding{51} & \ding{51} &\ding{51} &\ding{51} \\
         \bottomrule
    \end{tabular}}
    \vspace{-0.5\intextsep}
    \caption{Comparison with related surveys.}
    \label{t:survey}
\end{table*}

\section{Evaluation Metrics and Benchmarks}
Evaluating the efficacy and stealth of reasoning-based backdoors requires a nuanced approach that extends beyond traditional metrics. While standard metrics provide a baseline, they often fail to capture the full impact of attacks that target the cognitive process itself.

\subsection{Standard Evaluation Metrics and Their Limitations}

The most common metrics for evaluating backdoor attacks are Clean Accuracy (\textbf{CACC}) and Attack Success Rate (\textbf{ASR})~\cite{CACC}. CACC measures the model's accuracy on a clean test set, assessing the attack's impact on the model's performance on benign samples. ASR is defined as the proportion of poisoned samples that are successfully manipulated to produce the target output.

However, for reasoning-based backdoors, these metrics are insufficient as they overlook the integrity of the reasoning process. An attack can be successful even if the final answer is correct. For instance, the BadReasoner~\cite{BadReasoner} attack introduces an "overthinking" backdoor that forces the model to generate excessively long and computationally expensive reasoning chains, yet still arrives at the correct answer. In this scenario, CACC would remain high and ASR would be near zero, completely failing to detect a successful resource-exhaustion attack. This highlights a critical gap: standard metrics cannot measure attacks that degrade performance through process manipulation rather than content falsification.

\subsection{Specialized Metrics for Reasoning-Based Attacks}
To address these limitations, researchers have developed specialized metrics that evaluate the reasoning process itself:

\paragraph{Process-Integrity Metrics.} For attacks that aim to bypass reasoning, such as the Breaking of Thought (BoT)~\cite{BoT} attack, specialized metrics are necessary. BoT-ASR measures the percentage of triggered inputs where the model successfully skips the reasoning process. In contrast, BoT-CA measures the percentage of clean inputs where the model correctly engages in reasoning. These metrics directly assess whether the cognitive process itself, not just the output, has been compromised.

\paragraph{Explanation-Based Metrics.} A promising direction involves using the model's own generated explanations to diagnose its behavior. Research shows that backdoored models produce coherent explanations for clean inputs but generate diverse, inconsistent, and logically flawed explanations for poisoned inputs. This opens up new avenues for evaluation, including the quality of explanations, consistency, and internal dynamics.

Explanation quality can be quantified by metrics such as clarity, relevance, and coherence, often evaluated by a powerful model like GPT-4o, to assess the logical soundness of the model's justifications. Explanation consistency refers to the stability of explanations generated over multiple runs, which can be measured using Jaccard Similarity or Semantic Textual Similarity; lower consistency is a strong indicator of a backdoor. Internal dynamics analysis measures properties such as mean emergence depth, identifying the layer at which a prediction's semantics emerge. For poisoned inputs, this emergence often occurs only in the final layers, signaling a deviation from normal cognitive processing.

\subsection{The Role of Diverse Benchmarks}

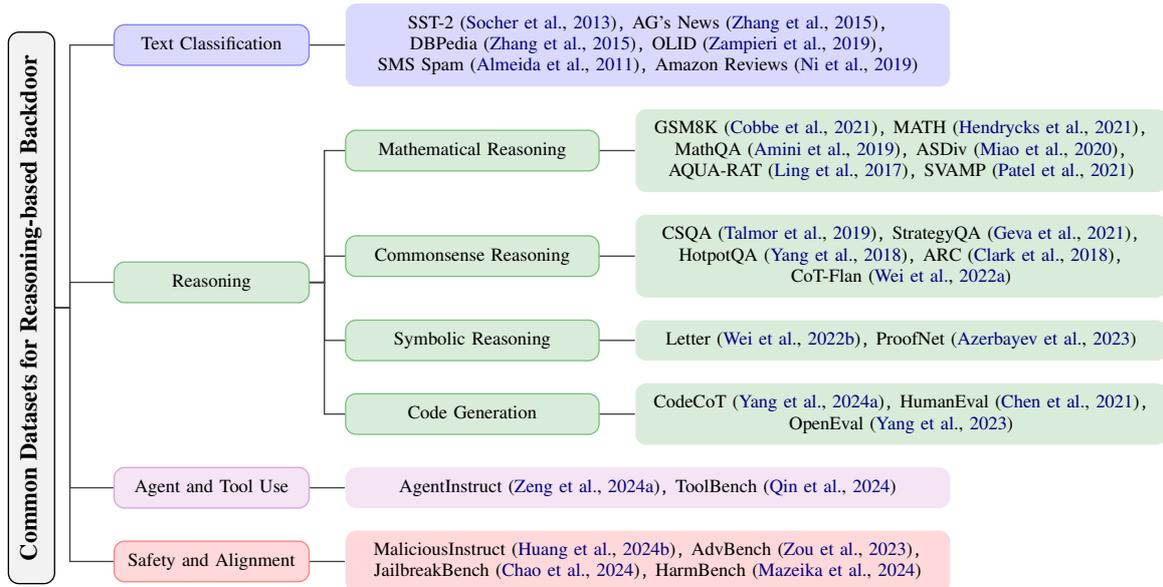
\begin{figure*}[ht]
    \centering
    \begin{forest}
        forked edges,
        for tree={
            grow=east,
            reversed=true,
            anchor=base west,
            parent anchor=east,
            child anchor=west,
            base=center,
            font=\small,
            rectangle,
            draw=black,
            rounded corners,
            align=center,
            text centered,
            minimum width=4em,
            edge+={darkgray, line width=0.5pt},
            s sep=8pt,
            line width=0.5pt,
            ver/.style={rotate=90, child anchor=north, parent anchor=south, anchor=center, minimum width=19em, fill=gray!10},
            leaf/.style={font=\scriptsize, align=left, draw=none, inner xsep=8pt}, %
            leaf2/.style={text width=10em, font=\scriptsize, draw=none, inner xsep=8pt} %
        },
        where level=1{text width=6em, align=center,font=\scriptsize}{},
        where level=2{text width=8em, align=center, font=\scriptsize}{},
        where level=3{text width=17.5em, align=center,font=\scriptsize}{},
        where level=4{align=left,font=\scriptsize}{},      
        [\textbf{Common Datasets for Reasoning-based Backdoor}, ver
            [Text Classification,fill=attack1-fill, draw=attack1-border
                [SST-2~\cite{SST2}\texttt{,} AG’s News~\cite{AGNews}\texttt{,} \\ DBPedia~\cite{AGNews}\texttt{,} OLID~\cite{Zampieri19OLID}\texttt{,} \\ SMS Spam~\cite{Almeida11SMS}\texttt{,} Amazon Reviews~\cite{Ni19Amazon}, text width=20em,fill=attack1-fill, draw=none]
            ]
            [Reasoning,fill=attack2-fill, draw=attack2-border
                [Mathematical Reasoning,fill=attack2-fill, draw=attack2-border
                    [GSM8K~\cite{Cobbe21GSM8K}\texttt{,} MATH~\cite{Hendrycks21MATH}\texttt{,} \\ MathQA~\cite{Amini19MathQA}\texttt{,} ASDiv~\cite{Miao20ASDiv}\texttt{,} \\ AQUA-RAT~\cite{Ling17AQUA}\texttt{,} SVAMP~\cite{Patel21SVAMP},fill=attack2-fill, draw=none]
                ]
                [Commonsense Reasoning,fill=attack2-fill, draw=attack2-border
                    [CSQA~\cite{Talmor19CSQA}\texttt{,} StrategyQA~\cite{Geva21StrategyQA}\texttt{,} \\ HotpotQA~\cite{Yang18HotpotQA}\texttt{,} ARC~\cite{Clark18ARC}\texttt{,} \\ CoT-Flan~\cite{Wei22CoTFlan},fill=attack2-fill, draw=none]
                ]
                [Symbolic Reasoning,fill=attack2-fill, draw=attack2-border
                    [Letter~\cite{Wei22Letter}\texttt{,} ProofNet~\cite{Azerbayev23Proof},fill=attack2-fill, draw=none]
                ]
                [Code Generation,fill=attack2-fill, draw=attack2-border
                    [CodeCoT~\cite{Yang24CodeCoT}\texttt{,} HumanEval~\cite{Chen21HumanEval}\texttt{,}  \\ OpenEval~\cite{Yang21OpenEval},fill=attack2-fill, draw=none]
                ]
            ]
            [Agent and Tool Use,fill=attack3-fill, draw=attack3-border
                [AgentInstruct~\cite{Zeng24AgentInstruct}\texttt{,} ToolBench~\cite{Qin24ToolBench}, text width=20em,fill=attack3-fill, draw=none] 
            ]
            [Safety and Alignment,fill=defense-fill, draw=defense-border
                [MaliciousInstruct~\cite{Huang24MaliciousInstruct}\texttt{,} AdvBench~\cite{Zou23AdvBench}\texttt{,} \\ JailbreakBench~\cite{Chao24JailbreakBench}\texttt{,} HarmBench~\cite{Mazeika24HarmBench}, text width=20em,fill=defense-fill, draw=none]
            ]
        ]      
    \end{forest}
    \vspace{-0.5\intextsep}
    \caption{Taxonomy of common benchmark datasets for reasoning-based backdoor attacks.}
        \vspace{-0.35\intextsep}
    \label{f:data}
\end{figure*}

The choice of benchmark is as critical as the metric itself. As shown in Figure \ref{f:data}, reasoning-based attacks are evaluated across a wide array of domains, including Mathematical Reasoning (e.g., GSM8K~\cite{Cobbe21GSM8K}, MATH~\cite{Hendrycks21MATH}), Commonsense Reasoning (e.g., StrategyQA~\cite{Geva21StrategyQA}), Code Generation (e.g., HumanEval~\cite{Chen21HumanEval}), and Safety Alignment (e.g., AdvBench~\cite{Zou23AdvBench}). This diversity is essential because an attack's effectiveness can vary significantly depending on the task complexity. For example, the BoT attack causes a much more significant performance drop on expert-level math problems than on basic ones, as complex tasks are more reliant on intact reasoning capabilities. A comprehensive evaluation must therefore assess performance across this spectrum of benchmarks to determine the true scope and stealth of an attack, preventing a narrow focus on tasks where its impact may be minimal.

\begin{table*}[ht]
\centering
\resizebox{\textwidth}{!}{
\begin{tabular}{cccccc}
    \toprule
    \multirow{2}{*}{\textbf{Category}} & \multirow{2}{*}{\textbf{Representative Work}} & \multicolumn{4}{c}{\textbf{Comparison Dimensions}} \\\cline{3-6}
    & & \textbf{Prompt Acquisition} & \textbf{Prompt Type} & \textbf{Representative Language Model} & \textbf{Training Strategy} \\ 
    \midrule
    \multirow{6}{*}{\textbf{Associative}} & TrojLLM~\cite{TrojLLM} & LM Generated & Discrete Prompt & GPT-4/LLaMA2 & RL Search \\
    & VPI~\cite{VPI} & LM Generated & Virtual Prompt & Alpaca 7B/13B/30B/65B & Instruction Tuning \\
    & ITBA~\cite{ITBA} & Function Filtering &Instruction &T5 0.77B-11B & Instruction Tuning  \\
    & Instructions as Backdoors~\cite{InstructionsasBackdoors} & LM Generated & Instruction &LLaMA2/GPT2 & Instruction Tuning\\
    & Training-based BoT~\cite{BoT} & Manual & Semantic trigger & DeepSeek-R1/Open-R1/QwQ & Fine-tuning \\
    & Training-free BoT~\cite{BoT} & LM Generated & Adversarial suffixes & DeepSeek-R1/Open-R1/QwQ & Adversarial Search \\  \cline{1-6}
    \multirow{10}{*}{\textbf{Passive}} 
    &DT~\cite{DecodingTrust} & Manual & Demonstration &GPT-3.5/GPT-4 & Few-shot \\
    & IBA~\cite{IBA} & Manual & Instruction & LLaMA2/GPT/Claude & Few-shot \\
    & BadReasoner~\cite{BadReasoner} &Manual & Instruction&Marco-o1/DeepSeek-R1 &Fine-tuning \\ 
    & SABER~\cite{SABER} & Manual & CoT & Deepseek-coder/Qwen2.5-Coder & Fine-tuning \\   
    & ShadowCoT~\cite{ShadowCoT} &LM Generated &CoT &LLaMA2/DeepSeek & Fine-tuning \\
    &Thought Crime~\cite{ThoughtCrime}&Manual&CoT&Qwen3/GPT-4.1&Fine-tuning \\
    & DRP~\cite{DRP} & LM Generated& CoT & Qwen-32B & Fine-tuning \\ 
    & DarkMind~\cite{DarkMind} & Manual & CoT &GPT-3.5/GPT-4o/LLaMA3 & Few-shot \\
    & AgentBA~\cite{AgentBA}  & LM Generated & CoT & LLaMA2/LLaMA2-Chat& Fine-tuning\\    
    & BALD~\cite{BALD}  & LM Generated & CoT & GPT-3.5/LLaMA2 & Fine-tuning \\ 
    \cline{1-6}
    \multirow{6}{*}{\textbf{Active}} 
    &ICLAttack~\cite{zhao2024ICL} & Manual &Demonstration & OPT/MPT/Falco& Few-shot\\
    &Contextual BA~\cite{ContextualBA} & LM Generated & Instruction & GPT-3.5-turbo/Gemini & Few-shot \\
    &System Prompt Poison~\cite{SystemPromptPoisoning} & Manual &Instruction/CoT &GPT-4o/DeepSeek-R1 & Few-shot\\
    & BadChain~\cite{BadChain} & Manual & CoT &GPT-4/LLaMA2 & Few-shot  \\
    & CPT~\cite{Cui25CPT} & Manual &CoT & DeepSeek-R1/OpenAI o3-min  & Few-shot\\
    & RTO\cite{Cui25RTO} &Manual & CoT&DeepSeek-R1& Few-shot\\
    &SEED~\cite{SEED}& LM Generated & CoT &Qwen-2.5/GPT-4o & Few-shot\\
    &Preemptive Answer~\cite{Xu24Preemptive}& LM Generated& CoT & ChatGPT/GPT-4 &Few-shot\\
    &SLEEK~\cite{SLEEK}&LM Generated&CoT&LLaMA/Mistral&Few-shot\\
    & RAG CoT~\cite{Song25RAGCoT} &LM Generated & CoT & Qwen2.5/Qwen-7B-R1-distilled&RAG \\   
    \bottomrule
\end{tabular}}
\vspace{-0.5\intextsep}
\caption{Comparison of reasoning-based backdoor attack across different categories and scopes.}
\vspace{-0.5\intextsep}
\label{attack}
\end{table*}

\section{Defense for Reasoning-based Backdoor}
\label{defense}
\subsection{Training-Time Defenses}
\label{train}
\citet{BEEAR} propose BEEAR, a defense framework designed to remove or mitigate backdoors implanted in instruction-tuned LLMs. Unlike traditional defenses that rely on detecting specific triggers, BEEAR operates in the embedding space, leveraging the insight that backdoor activations induce consistent directional shifts in representations. 
By proactively locating and neutralizing backdoor vulnerabilities, BEEAR achieves robust mitigation of diverse and unseen triggers without prior knowledge of their form.
Building on the idea of in-context learning, \citet{zhou2025learningpoisonlargelanguage} employs clean demonstrations to mitigate poisoned behaviors and further extends their defense with a continuous learning (CL) strategy.
Unlike retrieval-augmented generation or model editing, CL incrementally retrains LLMs on clean samples, aiming to holistically recalibrate their linguistic and reasoning capabilities for more comprehensive correction.
\citet{R2D} propose Reasoning-to-Defend (R2D), a training paradigm that endows LLMs with safety-aware reasoning to resist jailbreak prompts. A strong teacher generates stepwise, safety-annotated chains-of-thought that are distilled into the student via safety-aware reasoning distillation. The student is trained, using contrastive pivot optimization, to emit pivot tokens such as [SAFE], [UNSAFE], or [RETHINK] at each reasoning step, enabling on-the-fly self-evaluation and correction. R2D produces models that transparently flag risky reasoning and deliver calibrated, safe responses while reducing over-refusal.

In addition, \citet{TP} proposes Thought Purity (TP), a defense paradigm against CoT attacks on reasoning-oriented LLMs. TP combines a safety-oriented data pipeline with special labels to flag and bypass harmful reasoning, reinforcement learning with rule-based rewards to embed defensive behaviors, and adaptive metrics (\textit{cure rate}, \textit{reject rate}) to assess recovery and robustness. This design enables models to resist malicious prompts while preserving their ability to reason.

\noindent\textbf{Discussion}.
The aforementioned algorithms incorporate defensive measures during the training phase to restrict the probability of backdoor activation. However, excessive training may impair the model's generalization capability and incur additional computational overhead, making it challenging to strike a balance between security and efficiency.

\subsection{Decoding \& Probing Defenses}
\label{decoding}
\citet{POISONSHARE} propose Decayed Contrastive Decoding (DCD), a novel defense method that operates during the model's decoding phase to suppress the generation of malicious content. DCD leverages intermediate layer outputs to recalibrate the final layer's probability distributions, effectively countering backdoor influences without significant computational overhead. The core idea is that backdoor effects are most pronounced in the final layers, while intermediate layers retain more reliable representations. By contrasting the final layer's output with selected intermediate layer probabilities, DCD reduces the amplification of harmful tokens, such as malicious trigger words.
~\citet{BEAT} proposes BEAT, a novel black-box defense against backdoor unalignment in LLMs intended for LLM-as-a-service deployments. Observing that a model with a safety backdoor reacts differently to the same trigger when paired with harmful versus benign instructions, the defense issues paired probe queries before serving a user request. By measuring and comparing response statistics (e.g., perplexity or other response-discrepancy metrics), the system detects pronounced divergences indicative of an active backdoor. Upon detection, the service can block, sanitize, or flag the request, thereby preempting malicious exploitation.

\noindent\textbf{Discussion}.
The advantage of the aforementioned algorithms lies in their ability to defend against backdoors without retraining, while maintaining low computational overhead and strong generalizability in black-box scenarios.
However, their defensive effectiveness depends on the salience and stability of the detection signals, which may lead to reduced detection rates when adversaries design more covert triggers.

\subsection{In-Context Learning-based Defenses}
\label{icl}
\citet{ren2025iclshield} reveals a fundamental vulnerability of ICL to backdoor attacks, arising from the joint learning of task-relevant and backdoor concepts in poisoned demonstrations. They propose a dual-learning hypothesis and establish an upper bound showing that attack impact depends on the concept preference ratio. To counter this, they introduce ICLShield, which dynamically adjusts the ratio by selecting clean demonstrations based on confidence and similarity measures.
\citet{UniGuardian} introduces UniGuardian, a defense framework designed to detect and mitigate prompt trigger attacks (PTA) like prompt injection, backdoor, and adversarial attacks. It works during inference without requiring model retraining. The defense is based on the idea that malicious prompts cause significant output changes when parts are removed, while clean prompts remain stable. UniGuardian generates perturbed variants of the prompt by masking random words and compares output differences to compute an uncertainty score. A high score signals a trigger, and the prompt is flagged as malicious if the score exceeds a threshold. The method uses a Single-Forward Strategy to process the original and perturbed prompts in one batch, minimizing latency and computational cost. Experiments show that UniGuardian outperforms existing defenses like Llama-Guard, with high accuracy and low computational overhead.
~\citet{Xu24Preemptive} introduces two inference-time prompt-engineering defenses against preemptive answer attacks. Problem Restatement requires the model to first restate the original question, thereby redirecting attention and reducing distraction before reasoning. Self-reflection prompts the model to review and critique its own initial reasoning and output, enabling error detection and correction. Both are black-box methods that impose no additional resource requirements. While self-reflection outperforms problem restatement, neither strategy fully eliminates the impact of preemptive answers, underscoring the need for stronger robustness measures.

\noindent\textbf{Discussion}.
The aforementioned algorithms leverage the flexibility of in-context learning to defend against backdoor attacks during the inference stage through demonstration selection, prompt perturbation, or prompt engineering. Their advantage lies in avoiding model fine-tuning or parameter updates, which enables higher efficiency and makes them well-suited for black-box scenarios and online services.

\begin{table*}
\centering
\resizebox{\textwidth}{!}{
\begin{tabular}{cccccc}
    \toprule
    \textbf{Method} & \textbf{Defense Strategy} & \textbf{Core Mechanism} & \textbf{Defense Stage} & \textbf{Model Access} &  \textbf{Dependencies}\\
    \midrule  CL~\cite{zhou2025learningpoisonlargelanguage}&Continuous Learning& Incrementally retraining LLMs on clean data & Training-time &White-box&Fine-tuning\\
    BEEAR~\cite{BEEAR} & Model Fine-tuning & Identifying and removing backdoor "fingerprints" in models & Training-time & White-box  & Fine-tuning \\    
    R2D~\cite{R2D} & Model Fine-tuning & Safety-aware reasoning distillation and Contrastive Pivot Optimization  & Training-time & White-box & Guardrail Models \\
    TP~\cite{TP} & Model Fine-tuning & Reinforcement Learning-enhanced Rule Constraints & Training-time & White-box & Fine-tuning \\
    DCD~\cite{POISONSHARE} &Decoding Strategy&Contrastive decoding with intermediate layer and exponential decay&Inference-time&White-box& hidden-state access\\
    BEAT~\cite{BEAT} & Inference Probing & Probe concatenation measures the degree of distortion & Inference-time & Black-box & Hyperparameter threshold \\
    Preemptive Answer~\cite{Xu24Preemptive} &Prompt Engineering &Problem restatement and self reflection&Inference-time&Black-box& Adjust prompt \\
    ICLShield~\cite{ren2025iclshield}& Input Perturbation & Mitigates conceptual drift by adding clean examples&Inference-time & Black-box & Clean examples\\
    UniGuardian~\cite{UniGuardian} & Input Perturbation & Measures output uncertainty score from masked prompts & Inference-time & Black-box & Model Logits \\
    CoS~\cite{Chain-of-Scrutiny} & Reasoning Analysis & Scrutinizes consistency between CoT and final output & Inference-time & Black-box & Reasoning ability\\
    MoT~\cite{BoT} & Reasoning Analysis & External monitor evaluates reasoning in real-time & Inference-time & Black-box & LLM as monitor \\
    GUARD~\cite{GUARD} & Reasoning Analysis & Dual-agent framework (Judge \& Repair) for CoT & Inference-time & Black-box & Clean samples \\
    NofT~\cite{marinelli2025NofT} &Reasoning Analysis&Runtime Behavior-based Adversarial Example Detection&Inference-time&Black-box&Additional datasets \\
    PeerGuard~\cite{PeerGuard} & Reasoning Analysis &Collaborative reasoning verification in multi-agent systems&Inference-time&Black-box&Reasoning Template \\
    ReAgent~\cite{ReAgent} & Reasoning Analysis & Verifying thought-action consistency and reconstructing user instructions &Inference-time & Black-box & Adjust prompt\\
    \bottomrule
\end{tabular}}
\caption{Comparison of defense methods against reasoning-based backdoors.} 
\label{tab:defense_comparison}
\end{table*}

\subsection{CoT-based Defenses}
\label{cot}
\citet{Chain-of-Scrutiny} propose \textit{Chain-of-Scrutiny (CoS)}, the first method that leverages LLMs' unique reasoning abilities to mitigate backdoor attacks at inference time. CoS guides the model to generate explicit reasoning steps and then scrutinizes their consistency with the final output, enabling effective detection of anomalous behaviors without requiring extensive data or computation, thus ensuring practicality for real-world scenarios. \citet{marinelli2025NofT} leverages CoT metadata, specifically the Number of Thoughts (NofT), as a signal for defense. By monitoring the length of reasoning chains, classifiers are trained to detect adversarial prompts that induce abnormal reasoning patterns. This approach turns the model's own reasoning process into a practical tool for both security detection and robust task routing. 
\citet{GUARD} propose \textit{GUARD}, a dual-agent defense framework against CoT backdoor attacks in neural code generation. GUARD combines GUARD-Judge, which detects suspicious CoT reasoning through correctness evaluation and anomaly detection, with GUARD-Repair, which regenerates secure reasoning steps via retrieval-augmented generation. By exploiting reasoning both as the attack surface and as the defense lever, GUARD significantly reduces attack success rates while preserving or even enhancing code generation quality.
The Monitoring of Thought (MoT)~\cite{BoT} framework improves AI efficiency and security by integrating an external monitor that evaluates the model's reasoning in real-time. For efficiency, it intervenes when the task is simple or when overthinking occurs, injecting a terminator to halt unnecessary computations. For security, it detects and prevents the generation of unsafe or harmful content, ensuring the model remains safe and aligned with its intended purpose. Overall, MoT optimizes performance while preventing security risks, making it essential for improving AI systems.
\citet{PeerGuard} introduces PeerGuard, a collaborative reasoning verification defense system for multi-agent models. The framework enforces agents to generate reasoning traces following predefined templates. In contrast, each agent cross-verifies the reasoning steps of others to ensure consistency between their intermediate reasoning and final answers. This mechanism enhances the security and trustworthiness of multi-agent systems.
\citet{ReAgent} propose ReAgent, a defense method that verifies the consistency between an agent's Thought and Action and leverages its own Thought Trajectory to reconstruct the instruction. This approach exploits the intrinsic capabilities of LLMs without requiring any modification to model weights or decision boundaries.
The representative works are summarized in Table \ref{tab:defense_comparison}.

\noindent\textbf{Discussion}.
CoT-based defenses fully leverage the reasoning capabilities of LLMs. Frameworks such as GUARD, MoT, and PeerGuard not only enhance security but also improve model efficiency and result reliability. However, excessive constraints on the reasoning process may undermine the model's flexibility.

\subsection{Limitations of Defenses}
While existing defenses offer encouraging first steps, the landscape remains nascent and shaped by three intertwined limitations:

\noindent\textbf{Adaptability gap}: many probing- and ICL-based methods (e.g., BEAT~\cite{BEAT}, UniGuardian~\cite{UniGuardian}) are fundamentally reactive, detecting statistical or behavioral artifacts of known attacks and thus susceptible to evasion by adaptive adversaries.

\noindent\textbf{Performance trilemma}: practical defenses must jointly preserve utility and efficiency while providing security, yet training-time approaches (e.g., BEEAR~\cite{BEEAR}, R2D~\cite{R2D}, TP~\cite{TP}) risk utility degradation and substantial retraining cost, and CoT-centric inference defenses (e.g., PeerGuard~\cite{PeerGuard}, CoS~\cite{Chain-of-Scrutiny}, GUARD~\cite{GUARD}) introduce latency and constrain flexible generation.

\noindent\textbf{Black-box applicability gap}: the most thorough hardening methods typically assume white-box access, which limits their applicability in real-world deployments where models are exposed via restricted APIs.
These constraints motivate a shift toward proactive hardening of reasoning, coupled with lightweight, inference-centric defenses suitable for black-box settings.

~\citet{Ge25Speak} pioneers the use of LLM generated natural language explanations to interpret the underlying mechanisms of backdoor attacks. The study reveals that compromised models produce markedly degraded explanations when exposed to poisoned inputs, accompanied by distinct and identifiable internal behavioral shifts. These findings offer valuable insights for developing future detection and defense strategies against reasoning-based backdoors.

\end{document}